\documentclass[twocolumn,showpacs,preprintnumbers,amsmath,amssymb,superscriptaddress]{revtex4-1}

\usepackage{graphicx}
\usepackage{xcolor}
\usepackage{epsfig}
\usepackage{soul}
\usepackage{ulem}
\usepackage[colorlinks]{hyperref} 
\usepackage{cancel}

\begin{document}

\title{Spectral weight suppression in response functions of ultracold fermion-boson mixtures}
\author{Kai Ji}
\affiliation{Skolkovo Institute of Science and Technology, Nobel Street 3, 143026 Moscow Region, Russia}
\affiliation{TQC, Universiteit Antwerpen, Universiteitsplein 1, B-2610 Antwerpen, Belgium}
\affiliation{Institut f\"ur Theoretische Physik, Universit\"at Heidelberg, Philosophenweg 19, D-69120 Heidelberg, Germany}
\author{Andreas Komnik}
\email{komnik@uni-heidelberg.de}
\affiliation{Institut f\"ur Theoretische Physik, Universit\"at Heidelberg, Philosophenweg 19, D-69120 Heidelberg, Germany}

\date{\today}

\begin{abstract}
We study the dynamical response of ultracold fermion-boson mixture in the Bogoliubov regime, where the interactions between fermionic impurities and bosonic excitations (phonons) are described by an effective Fr\"{o}hlich model under the Bogoliubov approximation.
A characteristic suppression of the single-particle spectral weight is found in the small momentum region where the impurity band and phonon mode intersect.
Using diagrammatic technique we compute the Bragg spectra as well as the momentum dependent force-force correlation function. We find that both of them are heavily affected by the spectral weight suppression effect at low impurity densities in both 1D and 2D systems.
We show that the the spectral weight suppression feature in Bragg spectra, which was previously found in the quantum Monte Carlo simulations and which cannot be recovered by the random phase approximation, can be accurately reproduced with the help of vertex corrections.
\end{abstract}


\maketitle

\section{Introduction}
\label{sec:intro}

The experimental advances in manipulation of ultracold gas mixtures have made them outstanding platforms to explore the quantum many-body phenomena analogous to those in condensed-matter physics \cite{Feynman1982,Lloyd1996}.
In recent years, a high precision experimental control of such parameters as the strength of interaction, concentration of individual mixture components, imbalance of internal states, as well as dimensionality of  the system has become possible, thus allowing clean setups to study the properties of phase transitions, excitation spectra and dynamical processes \cite{Bloch2008,Giorgini2008,Chin2010}.
In the case of ultracold fermion-boson mixtures, the systems are realized by mixing atoms of different species \cite{Stan2004,Inouye2004,Schuster2012,Scelle2013} or isotopes of the same element \cite{Ferrier-Barbut2014}, in which e.~g. fermionic impurities are subject to interactions with the bosonic excitations (phonons) in Bose-Einstein condensate (BEC) upon sympathetic cooling \cite{Truscott2001,Modugno2002}.
In the Bogoliubov regime, where the depletion of condensate is negligible and the Bogoliubov linear approximation holds \cite{Pitaevskii2003}, such a complex system is nothing but an analog of the conventional Fr\"{o}hlich polaron in solid-state materials \cite{Cucchietti2006,Sacha2006,Tempere2009}.

The idea of polaron was first proposed by Landau and Pekar \cite{Landau1933,Pekar1946}, and further developed by Fr\"{o}hlich and Feynman \cite{Froehlich1950,Froehlich1954,Feynman1955} so as to describe the long-living states of quasi-particles comprising dressed electrons and collective excitations due to electron-phonon couplings in polar crystals.
This conceptual framework plays an important role in a number of intriguing quantum phenomena, such as e.~g. the conventional superconductivity (see Ref.~\cite{Alexandrov2009} for a review of the solid state polaron). 
There are, however, fundamental differences between the solid-state Fr\"{o}hlich polarons and their BEC counterparts.
The most obvious one is the different dispersion relation of the bosonic subsystem as well as profoundly different momentum dependence of the coupling strength \cite{Tempere2009}.
However, these details do not alter the general picture of the polaron and its quasi-particle nature and the Fr\"{o}hlich model hence serves as a reasonable starting point for interpreting the underlying intricacy of ultracold fermion-boson mixtures.

From another perspective, solving the Fr\"{o}hlich model is a nontrivial and challenging theoretical task.
While analytical approaches, such as Lee-Low-Pines variational ansatz \cite{Lee1953}, Landau-Pekar strong coupling theory \cite{Landau1948}, Feynman path integral and variational scheme \cite{Feynman1955} work well in many circumstances and especially in solid state systems, they are not able to yield a complete knowledge about the Fr\"{o}hlich polaron properties.
Recent studies on the self-trapping effect of BEC polaron in its ground state have exposed a marked discrepancy between the predictions of the Feynman's variational approach \cite{Casteels2012,Shashi2014} and the numerical results \cite{Vlietinck2015,Grusdt2015a,Shchadilova2014}.
While the analytical calculations predict a monotonic enhancement of the polaron binding energy with the increasing coupling strength, the numerical works predict a clear upper bound for the energy change.
Such a discrepancy was also noticed in a quantum Monte Carlo investigation \cite{PenaArdila2015}.
It has been proposed that an experimental setup of impurity-doped BEC may act as a benchmark for examining these theories \cite{Grusdt2015b}. 

Besides the ground state properties, understanding the dynamical properties of the BEC polaron is of great importance because many of them are experimentally accessible by virtue of existing ultracold spectroscopic techniques like radio frequency \cite{Chin2004,Gaebler2010,Feld2011,Kohstall2012,Koschorreck2012} and Bragg spectroscopy \cite{Stamper-Kurn1999,Steinhauer2002,Buechler2004,Ozeri2005,Ernst2010}, and hence enable us to gain further insights into the properties of the Fr\"{o}hlich model.
Earlier theoretical studies on the BEC polaron have identified characteristic phonon-induced structures in the Bragg spectra of impurities \cite{Casteels2011a,Casteels2011b,Ji2014}, indicative of polaron formation in the system.
In addition, quantum Monte Carlo (QMC) simulations revealed a highly nontrivial feature in the excitation spectra of such mixtures \cite{Ji2014} -- a characteristic suppression of spectral weight in response functions. 
However, the presence of this spectral weight suppression (SWS) was not corroborated in other analytical works and its origin remains a puzzle.
The motivation of present paper is to uncover the mechanism of the SWS and to clarify its influence on the fermion-boson mixture.
We shall show that SWS is a direct consequence of the fermion-phonon coupling.
Its properties can be adequately described by a Feynman diagrammatic calculation, which shows a rather good agreement with the QMC simulation results once the Fock term is taken into account.
This is done both for 1D and 2D systems.
Moreover, through an investigation on the momentum-dependent force-force correlation, which gives a direct access to a friction coefficient \cite{Petravic2008}, we find that at least at low impurity  densities their properties are considerably modified by the appearance of the SWS.

The paper is organized as follows:
In Sec.~\ref{sec:model} we introduce the effective Fr\"{o}hlich Hamiltonian of a fermion-boson mixture and the relevant physical quantities including the dynamical correlation and spectral functions.
In Sec.~\ref{sec:pseudo-gap}, we develop a diagrammatic formalism of dynamical response.
Then we discuss in detail the emergence of the SWS in connection with the numerical results for the spectra.
The higher order diagrammatic contributions from vertex corrections are evaluated and compared with QMC simulation results in Sec.~\ref{sec:vertex}.
Our conclusions is presented in Sec.~\ref{sec:conclusion}.

\section{Model and observables}

\label{sec:model}

We start with the effective Fr\"ohlich Hamiltonian for a BEC-fermion mixture  in the Bogoliubov regime (at low energies) \cite{Froehlich1954,Tempere2009},
\begin{eqnarray}
\label{H_perb} \label{eq:NRone}
H & = & H_0 + H_I \, ,
    \\
\label{H_perb0}
H_0 & = & \sum_{\bf k} (\epsilon_{\bf k} - \mu_I) \, a^{\dag}_{\bf k} a_{\bf k} + \sum_{\bf q} \omega_{\bf q} b_{\bf q}^{\dag} b_{\bf q} \, ,
    \\
\label{H_perb1}
H_I & = & {1 \over \sqrt{V}} \sum_{\bf k,q} V_{\bf q} \, a^{\dag}_{\bf q + k} a_{\bf k} \, \left( b_{\bf q} + b^{\dag}_{- \bf q} \right) \, .
\end{eqnarray}
Here $a^{\dag}_{\bf k}$ ($a_{\bf k}$) and $b^{\dag}_{\bf q}$ ($b_{\bf q}$) are the creation (annihilation) operators for a fermionic impurity of momentum $\bf k$ and a Bogoliubov phonon of momentum $\bf q$, respectively.
The fermionic impurity has a kinetic energy $\epsilon_{\bf k}$ [$\equiv k^2 / (2 m_I)$], a mass $m_I$, and a chemical potential $\mu_I$. 
\begin{eqnarray}
\label{phonondisp}   \nonumber
 \omega_{\bf q} = c_s |{\bf q}| \sqrt{1 + (\xi q)^2/2}
\end{eqnarray}
is the energy dispersion relation of the phonon mode, with $c_s = \left( \sqrt{2} m_B \xi \right)^{-1}$ the speed of sound in condensate, $\xi = 1 / \sqrt{8 \pi a_{BB} n_0}$ its healing length, $m_B$ the boson mass, $n_0$ the condensate density and $a_{BB}$ the boson-boson $s$-wave scattering length.
In Eq.~(\ref{H_perb1}), $V$ is the volume of the system
\footnote{In Ref.~\cite{Tempere2009}, the factor $V^{-1/2}$ in Eq.~(\ref{eq:NRone}) is absorbed in $V_{\bf q}$ and turns into $(2 \pi)^{-1/2}$ in the thermodynamic limit.}
and the fermion-phonon coupling constant is given by
\begin{eqnarray}
\label{couple1} 
V_{\bf q} = \lambda [(\xi q)^2/ ((\xi q)^2 + 2)]^{1/4} \, ,
\end{eqnarray}
where $\lambda = g_{\rm IB} \sqrt{n_0}$, with $g_{\rm IB}$ being the effective interaction strength between the impurities and Bogoliubov excitations, which can be adjusted by changing either the particle density or the $s$-wave scattering length in the impurity-boson collision processes.
If not explicitly stated, we shall use the polaronic units throughout our calculation.
That means the distance is measured in the units of $\xi$, time in the units of $m_I \xi^2/\hbar$ and energy in the units of $\hbar^2/(m_I \xi^2)$. 
In the numerical calculations, we consider a specific system of $^{6}$Li impurities in a BEC of $^{23}$Na which renders $m_B/m_I \approx 3.8$.

In this work, we are mainly interested in the dynamical response of the polarons.
While for the conventional solid state materials it can be probed by an optical absorption measurement, the method of choice for the cold atomic systems is Bragg spectroscopy \cite{Stamper-Kurn1999,Steinhauer2002,Buechler2004,Ozeri2005,Ernst2010}.
The measured spectra -- the impurity Bragg spectral function 
 [$\equiv \mathcal{R} ({\bf q}, \omega)$] -- is related to the (retarded) density-density correlation function [$\equiv \chi^R ({\bf q}, \omega)$] \cite{Pitaevskii2003,Casteels2011a},
\begin{eqnarray}
\label{eq:Bragg1}
\mathcal{R} ({\bf q}, \omega) = - {1 \over \pi} \mbox{Im} \, \chi^R ({\bf q}, \omega) \, .
\end{eqnarray}
Mathematically, one can first evaluate the correlation function $\chi ({\bf q}, i \omega_n)$ in the Matsubara representation, where $\omega_n = 2 n \pi / \beta$ is the Matsubara frequency and $\beta = 1 / (k_B T) $ is the inverse temperature (we set $k_B=1$ from now on).
The retarded correlation is then obtained by an analytic continuation, i.~e. imposing $i \omega_n \rightarrow \omega + i 0^+$.
In the Matsubara representation, the density-density correlation is expressed as
\begin{eqnarray}
\label{eq:Bragg2}
\chi ({\bf q}, i \omega_n) = - {1 \over V} \int^{\beta}_0 d \tau e^{i \omega_n \tau}
    \langle T_{\tau} \rho^{\dag} ({\bf q}, \tau) \rho ({\bf q}, 0) \rangle \, ,
\end{eqnarray}
where $\langle \cdots \rangle$ means ensemble average, $T_{\tau}$ is the imaginary time ordering operator, and
\begin{eqnarray}
\label{eq:fdensity}   \nonumber
\rho({\bf q})= \sum_{\bf k} a_{{\bf k} + {\bf q}}^{\dag} a_{\bf k} \, 
\end{eqnarray}
is the Fourier component of the fermion density operator.

While the Bragg spectrum is related to the {\it particle pair} correlation $\chi ({\bf q}, i \omega_n)$, there are, of course purely {\it single-particle} quantities. It is difficult to immediately observe them experimentally but they have very clear physical content and are easily accessible analytically. That is why we are also going to discuss them.
One of them is the single-particle fermion spectral function,
\begin{eqnarray}
\label{eq:fspec}
\mathcal{A} ({\bf k}, \varepsilon) = - {1 \over \pi} \mbox{Im} \, G^R ({\bf k}, \varepsilon) \, ,
\end{eqnarray}
and the other one is the phonon spectral function,
\begin{eqnarray}
\label{eq:bspec}
\mathcal{B} ({\bf q}, \omega) = - {1 \over \pi} \mbox{Im} \, D^R ({\bf q}, \omega) \, .
\end{eqnarray}
They are related to the impurity Matsubara Green's function,
\begin{eqnarray}             \label{fermionGFdefinition}
\mathcal{G} ({\bf k}, i \varepsilon_n) = - \int^{\beta}_0 d \tau e^{i \varepsilon_n \tau}
    \langle T_{\tau} a_{\bf k} (\tau) a^{\dag}_{\bf k} (0) \rangle \, ,
\end{eqnarray}
with $\varepsilon_n = (2n + 1) \pi / \beta, 2 n \pi / \beta$ the fermion/boson Matsubara frequency, and the phonon Matsubara function,
\begin{eqnarray}
\mathcal{D} ({\bf q}, i \omega_n) = - \int^{\beta}_0 d \tau e^{i \omega_n \tau}
    \langle T_{\tau} B_{\bf q} (\tau) B^{\dag}_{\bf q} (0) \rangle \, ,
\end{eqnarray}
where $B_{\bf q} = b_{\bf q} + b^{\dag}_{\bf -q}$, respectively \cite{Mahan2000}.

In the solid state systems there is yet another very important quantity, which gives access to the optical properties of the impurities. It is the force-force correlation function. In the present case of   uncharged impurities it can be directly related to the friction of the impurities while they are moving in the medium \cite{Petravic2008}. Right in the moment such setups move into the focal point of both experimenters as well as theorists (see e.~g. Ref.~\cite{Dasenbrook2013, Mathy:2012fk}), that is why we also analyze this particular correlation function.

In order to obtain a sensitive definition of the force-force correlation function we first take a look onto a drag force acting on impurities due to impurity-BEC interaction, which is given by (see Appendix~\ref{sec:app_force} for the derivation)
\begin{eqnarray}
\label{eq:force1}
{\bf F}_I = - {i \over \sqrt{V} } \sum_{{\bf k, q}} {\bf q} V_{\bf q} U_{\bf k,q} \, ,
\end{eqnarray}
where
\begin{eqnarray}
\label{eq:force2}
U_{\bf k,q} \equiv B_{\bf q} a^{\dag}_{\bf k} a_{\bf k-q}
\end{eqnarray} 
is a composite bosonic operator involving one bosonic -- the displacement operator -- $B_{\bf q}$, and two fermionic operators.
Here $\bf k$ and $\bf q$ denote the fermion and phonon momenta, respectively.
One can see from Eq.~(\ref{eq:force1}), that when the impurity emits or absorbs a phonon with momentum $\bf q$, the drag force exerted on the impurity is proportional to the coupling strength $V_{\bf q}$ and the phonon momentum $\bf q$.
To gain insight into the $\bf k$- and $\bf q$-dependence of the drag force, we introduce a momentum-dependent drag force ${\bf f}_{\bf k,q}$ in accordance with Eq.~(\ref{eq:force1}) (here we suppress the subscript `$I$' of the impurity and neglect the constant coefficient for simplicity),
\begin{eqnarray}
\label{eq:force3} \nonumber
{\bf f}_{\bf k,q} = - i {\bf q} V_{\bf q} U_{\bf k,q} \, ,
\end{eqnarray}
and we also define the {\it momentum-dependent force-force correlation} (MDFC) as
\begin{eqnarray}
\label{eq:ff1}                        \nonumber
\widetilde{C}_{\bf k,q}(t) &=& \langle T {\bf f}_{\bf k,q}(t) {\bf f}^{\dag}_{\bf k,q}(0) \rangle
    \nonumber \\
    &=& q^2 V_{\bf q}^2  \langle T U_{\bf k,q}(t) U^{\dag}_{\bf k,q}(0) \rangle \, ,
\end{eqnarray}
where $T$ is the time-ordering operator. One of its most important characteristics is its instantaneous value $t \to 0^+$, which is equal to an integral over all energies of its Fourier transform, 
\begin{eqnarray}
\label{eq:ff1}
C_{\bf k,q} = \lim_{t \to 0^+} \widetilde{C}_{\bf k,q}(t) \, , 
\end{eqnarray}
which we also refer to as MDFC by abuse of terminology. Later on we shall focus on this quantity instead of the full time-dependent MDFC. 
As mentioned above, the correlation function of total force  $\langle {\bf F}_I(t) {\bf F}_I^{\dag} (0)\rangle$ is related to the optical absorption and the current-current correlation of electrons in semiconductors \cite{Mahan2000}.
In the present work, since we are interested in the $\bf q$- and $\bf k$-component of the drag force, we shall calculate MDFC by the diagrammatic technique as mentioned above.

We first define a standard time-ordered three-particle Green's function as,
\begin{eqnarray}
\label{eq:ff2}                        \nonumber
K_{\bf k, q} (t) & = & - i \theta (t) \langle U_{\bf k,q} (t) U^{\dag}_{\bf k,q} \rangle
    \nonumber \\
& & - i \theta (-t) \langle U^{\dag}_{\bf k,q} U_{\bf k,q} (t) \rangle \, .
\end{eqnarray}
If $K_{\bf k, q} (t)$ is known, then MDFC is immediately obtained in the equal-time limit $t \rightarrow 0^+$.
It can be conveniently computed from the the Matsubara Green's function, defined according to
\begin{eqnarray}
\label{eq:ff3}                        \nonumber
{\mathcal K}_{\bf k, q} (\tau) & = & -\theta (\tau) \langle U_{\bf k,q} (\tau) U^{\dag}_{\bf k,q} \rangle
    \nonumber \\
& & - \theta (- \tau) \langle U^{\dag}_{\bf k,q} U_{\bf k,q} (\tau) \rangle \, .
\end{eqnarray}
Its Fourier transform is given by
\begin{eqnarray}
\label{eq:ff4}                        \nonumber
{\mathcal K}_{\bf k, q} (i \omega_n) = \int_0^\beta d \tau e^{i \omega_n \tau}
    \langle U_{\bf k,q} (\tau) U^{\dag}_{\bf k,q} \rangle \, .
\end{eqnarray}
In what follows we compute all of these correlations for a binary mixture of fermionic impurities immersed into a BEC.


\section{SWS in fermion-boson mixtures}
\label{sec:pseudo-gap}

\subsection{Diagrammatic approach to dynamical response}

\begin{figure}[htbp]
\centering
\includegraphics[width=0.45\textwidth]{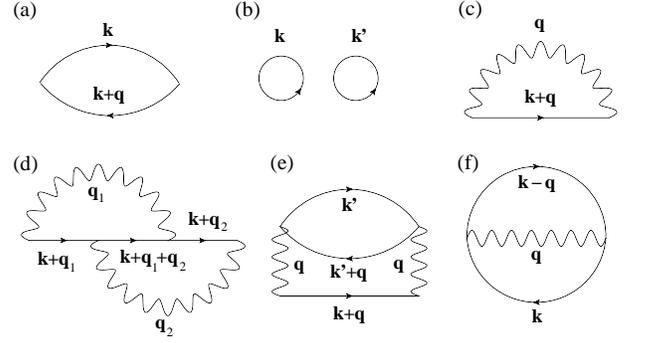}
\caption{Feynman diagrams for the interacting fermions (solid lines) and phonons (wavy lines).
(a) Loop diagram for the polarization of a pair of fermions and their density correlation.
(b) Self-correlation effect for a pair of fermions.
(c) Fock diagram for the first order self-energy of fermion.
(d) Vertex diagram for the second order fermion self-energy.
(e) Ring diagram for the second order fermion self-energy.
(f) Diagram for the force-force correlation.}
\label{fig:Diagram}
\end{figure}

The perturbative expansion of the density correlation in powers of interaction strength starts with the term of the zeroth order, which is obtained by replacing
 the ensemble average $\langle \cdots \rangle$ in Eq.~(\ref{eq:Bragg2}) by $\langle \cdots \rangle_0$ with respect to the eigenstate of the unperturbed Hamiltonian $H_0$.
Then one gets
\begin{eqnarray}
\label{eq:free1}
\chi ({\bf q}, i \omega_n) \simeq - {1 \over V} \int^{\beta}_0 d \tau e^{i \omega_n \tau}
    \langle T_{\tau} \rho^{\dag} ({\bf q}, \tau) \rho ({\bf q}, 0) \rangle_0 \, .
\end{eqnarray}
From the diagrammatic point of view Eq.~(\ref{eq:free1}) is a fermionic polarization loop [see the Feynman diagram in Fig.~\ref{fig:Diagram}(a)].
Using the Wick theorem, we obtain,
\begin{eqnarray}
\label{eq:free2}
\chi ({\bf q}, i \omega_n) & = & {1 \over \beta V} \sum_{{\bf k}, i \varepsilon_n} \mathcal{G}_0 ({\bf k}, i \varepsilon_n)
    \nonumber \\
& & \times 
\mathcal{G}_0 ({\bf k+q}, i \varepsilon_n + i \omega_n) ,
\end{eqnarray}
where $\mathcal{G}_0$ is the Matsubara Green's function (GF) of a free fermion as defined in \eqref{fermionGFdefinition}.
In Eq.~(\ref{eq:free2}), we have dropped a term for ${\bf q}=0$ as it corresponds to a time-independent self-correlation effect [see in Fig.~\ref{fig:Diagram}(b)], and is irrelevant to the dynamical response.
The fermion Matsubara GF  is related to the spectral function via
\begin{eqnarray}
\label{eq:free3}
\mathcal{G}_0 ({\bf k}, i \varepsilon_n) = \int d \varepsilon \, \frac{\mathcal{A}_0 ({\bf k}, \varepsilon)}{i \varepsilon_n - \varepsilon} ,
\end{eqnarray}
where $\mathcal{A}_0$ is the spectral function of a free fermion  as defined in Eq.~(\ref{eq:fspec}).
Substituting into Eq.~(\ref{eq:free2}) and following the standard procedure of frequency summation over $i \varepsilon_n$ \cite{Mahan2000}, one gets
\begin{eqnarray}
\label{eq:free4}
\chi ({\bf q}, i \omega_n) & = & {1 \over V} \sum_{\bf k}
    \int d \varepsilon d \varepsilon' \mathcal{A}_0 ({\bf k}, \varepsilon) \mathcal{A}_0 ({\bf k+q}, \varepsilon')
    \nonumber \\                        \nonumber
& & \times \frac{n_F (\varepsilon) - n_F (\varepsilon')}{i \omega_n + \varepsilon - \varepsilon'} .
\end{eqnarray}
According to Eq.~(\ref{eq:Bragg1}), the zeroth order Bragg spectral function can be determined after the analytic continuation,
\begin{eqnarray}
\label{eq:free5}
\mathcal{R} ({\bf q}, \omega) & = & - {1 \over \pi} \mbox{Im} \left[ \chi ({\bf q}, i \omega_n \rightarrow \omega + i 0^+) \right]
    \nonumber \\
& = & {1 \over V} \sum_{\bf k} \int d \varepsilon \, \mathcal{A}_0 ({\bf k}, \varepsilon) \mathcal{A}_0 ({\bf k+q}, \varepsilon + \omega)
    \nonumber \\
& & \times \left[ n_F (\varepsilon) - n_F (\varepsilon + \omega) \right] .
\end{eqnarray}
So it is a weighted convolution of two single-particle spectral functions.
Apparently, it characterizes an optical emission/absorption process in which a fermion is transferred from a state of momentum $\bf k$ and energy $\varepsilon$ to another one of momentum $\bf k+q$ and energy $\varepsilon + \omega$ with the assistance from a phonon of momentum $\bf q$ and energy $\omega$.
For the simplest case of a free fermion, the spectral function is
\begin{eqnarray}
\label{eq:free6}
\mathcal{A}_0 ({\bf k}, \varepsilon) = \delta [\varepsilon - (\epsilon_{\bf k} - \mu)] \, ,
\end{eqnarray}
so that Eq.~(\ref{eq:free5}) is simplified to
\begin{eqnarray}
\label{eq:free7}
\mathcal{R} ({\bf q}, \omega) & = & {1 \over V} \sum_{\bf k} \left[ n_F (\xi_{\bf k}) - n_F (\xi_{\bf k+q}) \right]
\nonumber \\
& & \times \delta (\omega + \xi_{\bf k} - \xi_{\bf k+q}) ,
\end{eqnarray}
where $\xi_{\bf k} = \epsilon_{\bf k} - \mu$.

We can perform the same program with the MDFC. Its Feynman diagram is shown in Fig.~\ref{fig:Diagram}(f).
In the lowest order approximation, assuming both bosons and fermions are undressed free particles, we find
\begin{eqnarray}
\label{eq:ff11}\label{eq:ff6}
& & {\mathcal K}_{\bf k, q} (i \omega_n) \simeq - \int_0^\beta d \tau e^{i \omega_n \tau}
    {\mathcal D}_0 ({\bf q}, \tau) {\mathcal G}_0 ({\bf k-q}, \tau) {\mathcal G}_0 ({\bf k}, - \tau)
    \nonumber \\
& & = \int d \varepsilon' d \varepsilon'' {\mathcal A}_0 ({\bf k-q}, \varepsilon') {\mathcal A}_0 ({\bf k}, \varepsilon'')
    \nonumber \\
& & \times \left\{ \frac{\left[ N_B (\varepsilon'' - \varepsilon') - N_B (\omega_{\bf q}) \right]
    \left[ n_F (\varepsilon') - n_F (\varepsilon'') \right]}
    {i \omega_n - \varepsilon' + \varepsilon'' - \omega_{\bf q}}
    \right. \nonumber \\
& & - \left. \frac{\left[ N_B (\omega_{\bf q}) - N_B (\varepsilon' - \varepsilon'')\right]
    \left[ n_F (\varepsilon') - n_F (\varepsilon'') \right]}
    {i \omega_n - \varepsilon' + \varepsilon'' + \omega_{\bf q}}
    \right\} \, .
\end{eqnarray}
To get Eq.~(\ref{eq:ff6}), we have used the spectral representation of free phonon propagator,
\begin{eqnarray}
\label{eq:freeph1}                        \nonumber
\mathcal{D}_0 ({\bf q}, i \omega_n) & = & \int d \omega \frac{\mathcal{B}_0 ({\bf q}, \omega)}{i \omega_n - \omega} ,
    \\
\label{eq:freeph2}                        \nonumber
\mathcal{B}_0 ({\bf q}, \omega) & = & \delta(\omega - \omega_{\bf q}) - \delta(\omega + \omega_{\bf q}) \, ,
\end{eqnarray}
as well as the result of Eq.~(\ref{eq:free3}) for the free fermion. Using Eq.~(\ref{eq:free6}) we can simplify it even further, and obtain
\begin{eqnarray}
& & {\mathcal K}_{\bf k, q} (i \omega_n) \simeq  \nonumber \\
& & = \frac{\left[ N_B (\xi_{\bf k} - \xi_{\bf k-q}) - N_B (\omega_{\bf q}) \right]
    \left[ n_F (\xi_{\bf k-q}) - n_F (\xi_{\bf k}) \right]}
    {i \omega_n + \xi_{\bf k} - \xi_{\bf k-q} - \omega_{\bf q}}
    \nonumber \\                        \nonumber
& & - \frac{\left[ N_B (\omega_{\bf q}) - N_B (\xi_{\bf k-q} - \xi_{\bf k}) \right]
    \left[ n_F (\xi_{\bf k-q}) - n_F (\xi_{\bf k}) \right]}
    {i \omega_n + \xi_{\bf k} - \xi_{\bf k-q} + \omega_{\bf q}} \, .
    \nonumber \\
\end{eqnarray}
After an analytic continuation, we obtain the spectral function of the force-force correlation from Eq.~(\ref{eq:ff6}),
\begin{eqnarray}
\label{eq:ff7}
\gamma_{\bf k,q} (\omega) & = & - {1 \over \pi} \mbox{Im} K^R_{\bf k,q} (\omega)
    \nonumber \\
& = & \delta \left( \omega + \xi_{\bf k} - \xi_{\bf k-q} - \omega_{\bf q} \right)
    \left[ n_F (\xi_{\bf k-q}) - n_F (\xi_{\bf k}) \right]
    \nonumber \\
& \times & \left[ N_B (\xi_{\bf k} - \xi_{\bf k-q}) - N_B (\omega_{\bf q}) \right]    
    \nonumber \\
& - & \delta \left( \omega + \xi_{\bf k} - \xi_{\bf k-q} + \omega_{\bf q} \right)
    \left[ n_F (\xi_{\bf k-q}) - n_F (\xi_{\bf k}) \right]
    \nonumber \\                        \nonumber
& \times & \left[ N_B (\omega_{\bf q}) - N_B (\xi_{\bf k-q} - \xi_{\bf k}) \right] \, ,
\end{eqnarray}
from which the greater and lesser GFs are recovered as
\begin{eqnarray}
\label{eq:ff8}
K^>_{\bf k,q} (\omega) & = & - i  2 \pi
\left[ N_B (\omega) + 1 \right] \gamma_{\bf k,q} (\omega) \, ,
    \\
\label{eq:ff9}
K^<_{\bf k,q} (\omega) & = & - i  2 \pi
N_B (\omega) \gamma_{\bf k,q} (\omega) \, .
\end{eqnarray}
Therefore, in the lowest order approximation, MDFC is found to be
\begin{eqnarray}
\label{eq:ff10}
& & C_{\bf k,q} = \lim_{t \rightarrow 0^+} i q^2 V^2_{\bf q} \int {d \omega \over 2 \pi} e^{-i \omega t} K^>_{\bf k,q} (\omega)
    \nonumber \\
& & = 
    q^2 V^2_{\bf q}
    \left[ n_F (\xi_{\bf k-q}) - n_F (\xi_{\bf k}) \right] \left\{ N_B (\xi_{\bf k} - \xi_{\bf k-q}) \right.
    \nonumber \\
& & 
    \times \left. \left[ N_B (\omega_{\bf q}) + 1 \right] 
    - N_B (\omega_{\bf q}) \left[ N_B (\xi_{\bf k-q} - \xi_{\bf k}) + 1 \right] \right\} \, .
\end{eqnarray}
Here one notices that the MDFC derived from $K^>_{\bf k,q} (\omega)$ in Eq.~(\ref{eq:ff8}) describes the force-force correlation of a fermion being scattered out of the Fermi sea by a phonon.
Analogously, there is another MDFC connected with the lesser GF $K^<_{\bf k,q} (\omega)$ in Eq.~(\ref{eq:ff9}), which corresponds to a hole scattered by a phonon.
Since the hole state is beyond our interest, we do not discuss it.

Needless to say, the zero-order result Eq.~(\ref{eq:free5}) is only a function of Fermi-Dirac distribution $n_F (\varepsilon)$ and does not contain any information about the polaron effect.  So does Eq.~\eqref{eq:ff10}.
In an earlier work \cite{Ji2014}, we have studied the fermion-phonon correlation in Bragg spectra by a random phase approximation (RPA).
That formalism has an advantage to reveal the interaction effects in a simple way.
Although in that calculation several polaron-related features have been identified in the Bragg spectra, the SWS could not be seen.
In the present work, we shall investigate these effects by employing a different strategy.
 For the purpose of highlighting the differences and similarities of both approaches, we provide a brief discussion of RPA in Appendix~\ref{sec:app_rpa}.

In order to capture the essence of the SWS, we make one step further and look into the contribution from the terms of higher order in fermion-boson interaction. 
One way to do that is to replace $\mathcal{G}_0$ in Eqs.~(\ref{eq:free2}) and (\ref{eq:ff6}) with the GF of a dressed fermion $\mathcal{G}$ \cite{Mahan2000}, which we assume to take the form 
\begin{eqnarray}
\label{eq:hf1}
\mathcal{G} ({\bf k},i \varepsilon_n) = \frac{1}{i \varepsilon_n -\xi_{\bf k} -\Sigma ({\bf k}, i \varepsilon_n)} ,
\end{eqnarray}
where $\Sigma ({\bf k}, i \varepsilon_n)$ is the self-energy or memory function of a dressed fermion. 
Inspired by Eq.~(\ref{eq:free5}), the Bragg spectral function can be expressed as
\begin{eqnarray}
\label{eq:hf2}
\mathcal{R} ({\bf q}, \omega) & = & {1 \over V} \sum_{\bf k} \int d \varepsilon \mathcal{A} ({\bf k}, \varepsilon) \mathcal{A} ({\bf k+q}, \varepsilon + \omega)
    \nonumber \\
& & \times \left[ n_F (\varepsilon) - n_F (\varepsilon + \omega) \right] ,
\end{eqnarray}
with $\mathcal{A} ({\bf k}, \varepsilon)$ being the single-particle spectral function of a {\it dressed} fermion. The same procedure applied to the MDFC leads to
\begin{eqnarray}
\label{eq:ff11}
& & {\mathcal K}_{\bf k, q} (i \omega_n) \simeq - \int_0^\beta d \tau e^{i \omega_n \tau}
    {\mathcal D}_0 ({\bf q}, \tau) {\mathcal G} ({\bf k-q}, \tau) {\mathcal G} ({\bf k}, - \tau)
    \nonumber \\
& & = \int d \varepsilon' d \varepsilon'' {\mathcal A} ({\bf k-q}, \varepsilon') {\mathcal A} ({\bf k}, \varepsilon'')
    \nonumber \\
& & \times \left\{ \frac{\left[ N_B (\varepsilon'' - \varepsilon') - N_B (\omega_{\bf q}) \right]
    \left[ n_F (\varepsilon') - n_F (\varepsilon'') \right]}
    {i \omega_n - \varepsilon' + \varepsilon'' - \omega_{\bf q}}
    \right. \nonumber \\
& & - \left. \frac{\left[ N_B (\omega_{\bf q}) - N_B (\varepsilon' - \varepsilon'')\right]
    \left[ n_F (\varepsilon') - n_F (\varepsilon'') \right]}
    {i \omega_n - \varepsilon' + \varepsilon'' + \omega_{\bf q}}
    \right\} \, ,
\end{eqnarray}
and 
\begin{eqnarray}
\label{eq:ff13}
C_{\bf k,q} & = & 
    q^2 V^2_{\bf q}
    \int d \varepsilon d \omega {\mathcal A} ({\bf k}, \varepsilon)
    \nonumber \\
& & 
    \times \left\{ {\mathcal A} ({\bf k-q}, \varepsilon + \omega - \omega_{\bf q})
    \left[ N_B (\omega_{\bf q}) + 1 \right]
    \right. \nonumber \\
& &     \times N_B (\omega_{\bf q} - \omega)
    \left[ n_F (\varepsilon + \omega - \omega_{\bf q}) - n_F (\varepsilon) \right]
    \nonumber \\
& &
    - {\mathcal A} ({\bf k-q}, \varepsilon + \omega + \omega_{\bf q})
    \left[ N_B (\omega_{\bf q} + \omega) + 1 \right]
    \nonumber \\
& &     \times \left. N_B (\omega_{\bf q})
    \left[ n_F (\varepsilon + \omega + \omega_{\bf q}) - n_F (\varepsilon) \right]
    \right\} \, .
\end{eqnarray}

The last expressions contain
the Fermi-Dirac distribution $n_F (\varepsilon)$, the Bose-Einstein distribution $N_B (\varepsilon)$ and the fermionic spectral function $\mathcal{A} ({\bf k}, \varepsilon)$.
The most important features of the polaron effect are captured in the spectral function $\mathcal{A} ({\bf k}, \varepsilon)$ through the fermion self-energy \footnote{We remark, that Eq.~\eqref{eq:hf2} does not represent an exact result even in the case the single-particle spectral function is known. We are, however, consistent in using it within the approximation scheme we employ.}.
In terms of Eq.~(\ref{eq:hf1}), $\mathcal{A} ({\bf k}, \varepsilon)$ can be written as a function of the retarded fermion self-energy $\Sigma^R ({\bf k}, \varepsilon)$ [$\equiv \mbox{Re} \Sigma^R ({\bf k}, \varepsilon) + i \mbox{Im}  \Sigma^R ({\bf k}, \varepsilon)$],
\begin{eqnarray}
\label{eq:hf3}
\mathcal{A} ({\bf k}, \varepsilon) = - {1 \over \pi}
    \frac{\mbox{Im} \Sigma^R ({\bf k}, \varepsilon)}{\left[ \varepsilon - \xi_{\bf k} - \mbox{Re} \Sigma^R ({\bf k}, \varepsilon) \right]^2 + \left[\mbox{Im}  \Sigma^R ({\bf k}, \varepsilon) \right]^2} \, .
    \nonumber \\
\end{eqnarray}
If the fermion self-energy is known, then the spectral function $\mathcal{A} ({\bf k}, \varepsilon)$ and Bragg spectral function $\mathcal{R} ({\bf q}, \omega)$ in Eq.~(\ref{eq:hf2}) as well as $C_{\bf k,q}$ from Eq.~\eqref{eq:ff13} can be determined immediately. It is important to realize that although $\mbox{Im}  \Sigma^R ({\bf k}, \varepsilon)$ can become very small, it is never zero in real systems. The reason is that this quantity even in non-interacting systems is proportional to reciprocal lifetime, or energy level width, of the corresponding state (e.~g. due to coupling to thermal reservoirs). In order to ensure non-vanishing spectral functions during numerical evaluations we replace $\mbox{Im}  \Sigma^R ({\bf k}, \varepsilon)$ by a numerical infinitesimal whenever it becomes too small. Its precise numerical values are given further down.

In order to obtain the self-energy we perform a diagrammatic calculation.
Fig.~\ref{fig:Diagram}(c) shows the irreducible diagram for the lowest order fermion self-energy,  coming only from the Fock term.
Here we use the free fermion propagator $\mathcal{G}_0$ and free phonon propagator $\mathcal{D}_0$.
The frequency summation is readily done and one gets
\begin{eqnarray}
\label{eq:hf4}
\Sigma^{(1)}  ({\bf k},  i \varepsilon_n)
& = & {1 \over V} \sum_{\bf q} V_{\bf q}^2
    \left[ \frac{N (\omega_{\bf q}) + n_F (\xi_{\bf k+q})}{i \varepsilon_n + \omega_{\bf q} - \xi_{\bf k+q}} \right.
    \nonumber \\
& & \left. + \frac{N (\omega_{\bf q}) + 1 - n_F (\xi_{\bf k+q})}{i \varepsilon_n - \omega_{\bf q} - \xi_{\bf k+q}} \right] ,
\end{eqnarray}
where $N (\omega_{\bf q}) = 1 / \left[ \exp (\beta \omega_{\bf q}) - 1 \right]$ and $n_F (\xi_{\bf k}) = 1 / \left[ \exp (\beta \xi_{\bf k}) + 1 \right]$ are the bosonic and fermionic distributions, respectively.
Here the superscript `(1)' reflects the fact that only the single phonon process is considered in the fermion self-energy. Obviously, $\Sigma^{(1)}$ ($\propto V_{\bf q}^2$) is the result of a standard second order perturbation calculation.
The next order contributions from the two-phonon processes are presented in Figs.~\ref{fig:Diagram}(d) and ~\ref{fig:Diagram}(e).
They correspond to the vertex and ring corrections, respectively. We discuss them in Sec.~\ref{sec:vertex}.

Eq.~(\ref{eq:hf4}) can be rewritten as
\begin{eqnarray}
\label{eq:hf6}
\Sigma^{(1)} (i \varepsilon_n) = \sum_j {C_j \over i \varepsilon_n + E_j} \, ,
\end{eqnarray}
where $C_j$ and $E_j$ represent some well-defined quantities.
After the analytic continuation $i \varepsilon_n \rightarrow \varepsilon + i 0^+$, the imaginary part of retarded self-energy can be found to be given by
\begin{eqnarray}
\label{eq:hf7}
\mbox{Im} \Sigma^{(1) R} (\varepsilon) = - \pi \sum_j C_j \delta (\varepsilon + E_j) \, ,
\end{eqnarray}
where $\delta (\varepsilon)$ is the Dirac delta function.
Substituting it into Eq.~(\ref{eq:hf3}) and taking into account that $C_j$ are non-negative, one immediately confirms the positivity of the spectral function $\mathcal{A} ({\bf k}, \varepsilon)$  within this approximation.
In the light of Eq.~(\ref{eq:hf3}), $\mathcal{A} ({\bf k}, \varepsilon)$ turns out to be a positive-definite Lorentzian located at $\varepsilon = \xi_{\bf k} + \mbox{Re} \Sigma^{(1) R} ({\bf k}, \varepsilon)$ and with a half-width-at-half-maximum of $-\mbox{Im} \Sigma^{(1) R} ({\bf k}, \varepsilon)$.
If the self-energy vanishes, for example in the non-interacting limit $\lambda \rightarrow 0$, the Lorentzian is reduced to a delta function.
In this case, the free fermion spectral function Eq.~(\ref{eq:free6}) is recovered (for the numerical details see the remark above).

\subsection{Emergence of the SWS: lowest order self-energy approximation}

 SWS can already be seen in the lowest order self-energy approximation presented above. First we turn to
 the numerical results on the spectral function $\mathcal{A} ({\bf k}, \varepsilon)$ and Bragg spectral function $\mathcal{R} ({\bf q}, \omega)$. In order to clarify the physical origin of the  SWS, we concentrate on a 1D system.
Then the fermion and phonon momenta are scalars.
Some results on 2D will be provided later in Sec.~\ref{sec:vertex}, and the mechanism discussed in this section can be easily generalized to the higher dimensions.
In our numerical calculations, we employ a cutoff for the fermion momentum in Eq.~(\ref{eq:hf2}) as $-\pi \le k \le \pi$,  and the phonon momentum in Eq.~(\ref{eq:hf4}) as $-2 \pi \le q \le 2 \pi$.
As demonstrated below, the SWS takes place in the small momentum region, $\lesssim 0.2 \pi$, thus the cutoff effect on the SWS can be safely disregarded. This assumption was confirmed by an exemplary calculation for a considerably larger cutoff for one typical constellation of other parameters (see Fig.~\ref{fig:DOS} below).

\begin{figure}[htbp]
\includegraphics[width=0.5\textwidth]{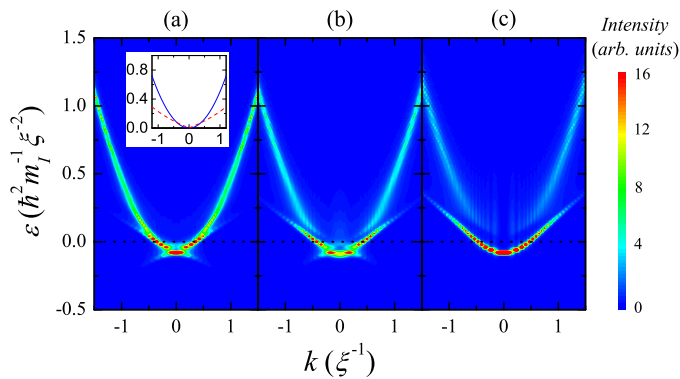}
\caption{(Color online) Emergent avoided crossing in the spectral functions of fermionic impurities in 1D BEC with increasing interaction strength: (a) $\lambda = 0.3$, (b) $\lambda = 0.4$, and (c) $\lambda = 0.5$.
The inset in (a) shows the dispersion relations of a free fermion (blue solid curve) and the Bogoliubov mode (red dashed curve).
The Fermi momentum is at $k_F = 0.1 \pi \xi^{-1}$.
The horizontal black dotted line denotes the Fermi level.
The temperature is set to be $T = 0$.}
\label{fig:A}
\end{figure}

Fig.~\ref{fig:A} shows the intensity graphs of the fermion spectral function $\mathcal{A} (k, \varepsilon)$ at temperature $T = 0$ for three different interaction strengths.
With the increase of $\lambda$, one clearly observes a band repulsion in the small $k$ region, leading to an energy gap between the upper and lower branches. 
In the inset of panel (a), we show the dispersion relations of the free fermion and Bogoliubov mode.
Comparing these energy dispersions with the main graphs, one can see, that the energy gap corresponds to an avoided crossing between the free fermion band and the Bogoliubov phonon mode. Being absent in a free system, the gap intensifies for growing interaction strengths. 

The physical reason for this effect is quite lucid. The ordinary single-particle spectral function describes, how easy it is to move a single particle with the fixed energy into the continuum outside of the system. Alternatively one can think about a probability to succeed in extracting a certain amount of energy. In a free system the shape of this probability distribution obviously follows the spectral function of the constituent fermions. In an interacting system two kinds of excitations in a system couple to each other. At the point, where their dispersions cross, they compete and it gets increasingly difficult to `take out' a fermion out of the system. That is why we see an avoided crossing shown in Fig.~\ref{fig:A}.

\begin{figure}[thbp]
\includegraphics[width=0.4\textwidth]{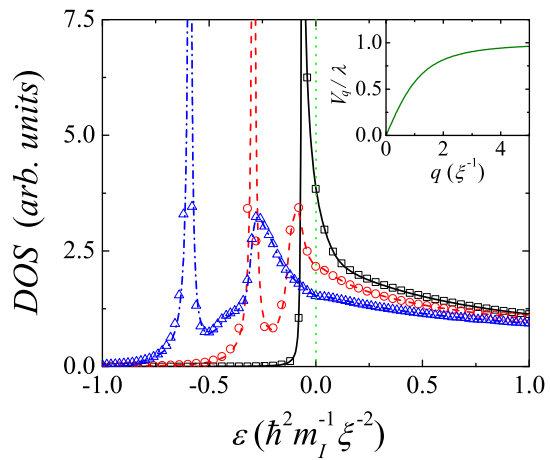}
\caption{(Color online)
Density of states for fermions in 1D BEC when $k_F = 0.1 \pi \xi^{-1}$ (black squares and solid curve), $0.2 \pi \xi^{-1}$ (red circles and dashed curve), and $0.3 \pi \xi^{-1}$ (blue triangles and dash-dotted curve), respectively.
The coupling constant is $\lambda = 0.5$.
The vertical green dotted line denotes the Fermi level.
The symbols are calculated with a momentum cutoff at $2 \pi \xi^{-1}$, and the curves are for a cutoff at $8 \pi \xi^{-1}$.
The inset shows the $q$-dependence of the ratio $V_q / \lambda$.
}
\label{fig:DOS}
\end{figure}

 In Fig.~\ref{fig:DOS} we plot the total density of states (DOS) for the fermions, which is obtained from the spectral function $\mathcal{A} (k, \varepsilon)$ by an integration over $k$.
Although the SWS can already be seen in Fig.~\ref{fig:A} at $\lambda = 0.5$, it can hardly be observed in Fig.~\ref{fig:DOS} when $k_F = 0.1 \pi$.
The structure becomes visible only when $k_F$ is large enough.
This behavior is due to the momentum dependence of the effective coupling strength $V_{\bf q}$.
When $k_F$ is small, the fermions can be excited out of Fermi sea by interacting with a phonon of small momentum $q$, which is associated with a rather weak coupling $V_{\bf q}$ according to Eq.~(\ref{couple1}) (see also in the inset of Fig.~\ref{fig:DOS}).
In this case, a narrow SWS exists, but it  is superimposed by the van Hove singularity in its vicinity.
With the increase of $k_F$, the momentum of the scattered phonon and the effective coupling increase quickly. As a result, the SWS also grows with $k_F$.

In order to make sure the growth of SWS with $k_F$ is irrelevant to the momentum cutoff (note that $k_F = 0.3 \pi$ is already comparable to the inverse healing length and cannot be regarded as a small wave vector), we raise the upper bound of $q$ from $2 \pi$ to $8 \pi$ in calculating the self-energy of Eq.~(\ref{eq:hf4}).
The results of two different upper bounds are presented in Fig.~\ref{fig:DOS}.
The good agreement between them indicates that when $k_F \sim 1$, as far as the  SWS is concerned, the cutoff at $2 \pi$ does not produce significant errors.

As mentioned above, the SWS comes from the avoided crossing between the fermion band and Bogoliubov mode.
This can be clearly seen in Fig.~\ref{fig:DOS}, where the location of the SWS points to the repulsion region between the two bands.
This property is different from that of conventional energy gaps.
For example, the phonon-induced Peierls gap is known to shift with the Fermi level.
The onset of the Peierls gap also indicates a metal-insulator phase transition in solid state materials.
In the fermion-boson mixture the SWS comes from the renormalization of fermion energy dispersion. Therefore no phase transition is incurred with the gap opening.

\begin{figure}[thbp]
\includegraphics[width=0.5\textwidth]{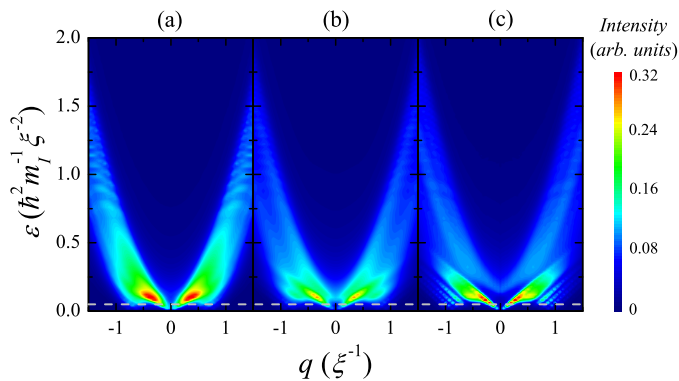}
\caption{(Color online) Avoided crossing seen in the Bragg spectroscopy of fermions in 1D BEC with increasing interaction strength: (a) $\lambda = 0.3$, (b) $\lambda = 0.4$, and (c) $\lambda = 0.5$.
The other parameters are the same as in Fig.~\ref{fig:A}.
The horizontal gray dashed line denotes the Fermi energy.
}
\label{fig:Bragg}
\end{figure}

According to Eq.~\eqref{eq:hf2} the Bragg spectrum can be written down in terms of the single-particle spectral function and thus should show the SWS feature too.
In Fig.~\ref{fig:Bragg}, we plot the respective intensity images of $\mathcal{R} (q, \omega)$, using the same parameters as in Fig.~\ref{fig:A}.
As in Fig.~\ref{fig:A}, the Bragg spectra also display a band repulsion for growing interaction strength $\lambda$.
Especially in panel (c) the upper and lower branches can be clearly distinguished.

In order to acquire an intuitive picture on the mechanism of SWS in Bragg spectrum, let us assume that the system contains only one fermion of momentum $k_0 \ll 1$ with energy
 $\varepsilon_0 = \hbar^2 k_0^2 / (2 m_I)$.
Its spectral function can be approximated as
$\mathcal{A} (k, \varepsilon) = \delta_{k,k_0} \delta ( \varepsilon - \varepsilon_0)$.
Putting this spectral function into Eq.~(\ref{eq:hf2}), one gets
\begin{eqnarray}
\label{eq:hf5}
\mathcal{R} (q, \omega) 
& = & 
{1 \over V} \mathcal{A} (k_0+q, \varepsilon_0+\omega) \left[ n_F(\varepsilon_0) - n_F(\varepsilon_0+\omega) \right]
\nonumber \\
& &
\rightarrow {1 \over V} \mathcal{A} (k_0+q, \varepsilon_0+\omega) ,
\end{eqnarray}
where $n_F (\varepsilon_0)=1$ and $n_F (\varepsilon_0+\omega)=0$ are for the initially occupied and unoccupied states, respectively.
Eq.~(\ref{eq:hf5}) shows that the Bragg spectrum can be directly mapped on the fermion spectral function, which maps out all the occupied states seen by this single fermion.
Although this scenario holds strictly for the system of a single fermion only, it is also approximately valid for systems of low impurity density.
This is the reason why SWS shows up in Fig.~\ref{fig:Bragg}(c).
However, if there are more fermions in the system, then the superposition of different fermion states in the Bragg spectrum must reduce its resemblance to the one-fermion spectral function and eventually the SWS gets less pronounced. 
(Results on Bragg spectra at higher Fermi levels show almost no dependence on the SWS and that is why we do not discuss them here.)
In spite of this, we can say that the SWS is rather distinctive at low impurity density.
In addition, Eq.~(\ref{eq:hf5}) also suggests that the Bragg spectroscopy might be applied as a technique to measure the fermion spectral function comparable to the radio frequency experiment  if the impurities are dilute \cite{Chin2004,Gaebler2010,Feld2011,Kohstall2012,Koschorreck2012}.

\begin{figure}[thbp]
\includegraphics[width=0.45\textwidth]{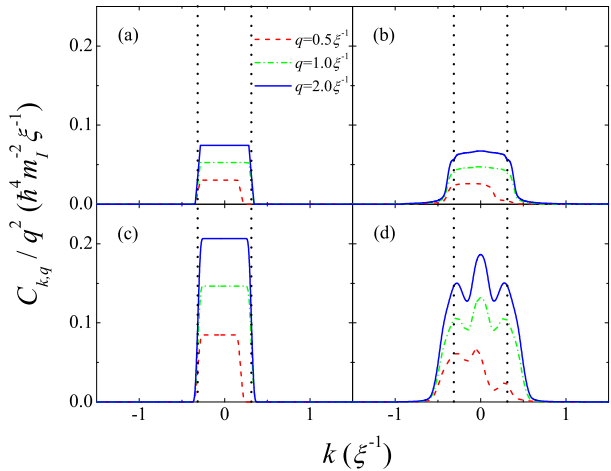}
\caption{(Color online) Momentum dependence of force-force correlation weights $C_{k,q}$,  divided by $q^2$ at a low fermion density with $k_F = 0.1 \pi \xi^{-1}$ at zero temperature.
The vertical black dotted lines illustrate $\pm k_F$.
Panels (a) and (b) correspond to $\lambda =0.3$.
Panels (c) and (d) are for $\lambda = 0.5$.
The free fermion spectral function is used in calculating (a) and (c), while the data in (b) and (d)  are computed with the help of the dressed fermion spectral function. 
}
\label{fig:Low}
\end{figure}

Now we turn to the MDFC. Fig.~\ref{fig:Low} presents a case of low fermion concentration, when the Fermi momentum is set at $k_F = 0.1 \pi$.
The upper panels (a) and (b) are for $\lambda = 0.3$, and the lower two are for $\lambda = 0.5$.
Panels (a) and (c) are calculated using Eq.~(\ref{eq:ff10}) under the free fermion assumption, while panels (b) and (d) are computed with the help of Eq.~\eqref{eq:ff11} for a dressed fermion, upon replacement of the free spectral function ${\cal A}_0$ by ${\cal A}$, computed in the lowest order self-energy approximation.
In each panel, the MDFC is plotted as a function of the fermion momentum $k$ for three different phonon momenta.

One feature of MDFC is that its carrier in the momentum space is roughly confined in a window of width $2 k_F$.
For example, in Fig.~\ref{fig:Low}(a), MDFC for $q = 1.0$ and $2.0$ shows non-zero values only for $-k_F \le k \le k_F$.
This feature arises from the fact that the fermion-phonon scatterings are allowed only inside the window for all states occupied by fermions.
On the contrary, outside of the window fermion states are empty and hence no scattering can take place.
The situation becomes somewhat different for $q = 0.5$, where the width of MDFC is smaller than $2 k_F$, and a small dent appears between the right edge of MDFC and $k_F$.
This suppression of MDFC is because $q = 0.5$ is too small to kick a fermion out of the Fermi sea from this region, i.~e. $k-q < -k_F$ cannot be satisfied.
Thus we conclude that the MDFC can be interpreted as a probability distribution of the fermion-phonon scattering process efficiency in the momentum space.

Panel (b) shows MDFC calculated by Eq.~(\ref{eq:ff13}) including the effect of Fock self-energy.
It is slightly broadened in comparison to the non-interacting plot (a) due to interactions.
As the coupling grows the MDFC spectrum computed with the `dressed' GF changes dramatically, developing a trident-like structure at large $q$, see panel (d) in Fig.~\ref{fig:Low}.
This is to be contrasted with the data in panel (c), where the free fermion GF is used.
This transition from a rectangular to trident-shaped feature occurs due to a strong renormalization of the fermion energy dispersion near the avoided crossing region, a remarkable manifestation of the SWS.

\begin{figure}[thbp]
\includegraphics[width=0.45\textwidth]{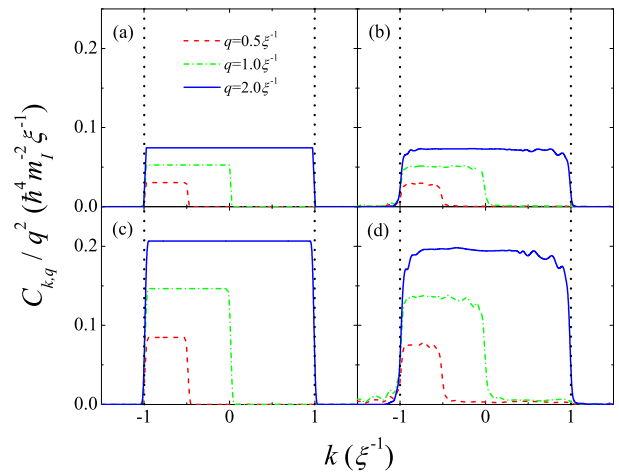}
\caption{(Color online) The same as in Fig.~\ref{fig:Low} for a higher impurity density with $k_F = \xi^{-1}$. 
}
\label{fig:High}
\end{figure}

To ensure the above modification of MDFC is indeed related to the SWS, we plot the same quantities at higher fermion density in Fig.~\ref{fig:High}.
Here the Fermi momentum is $k_F = 1$ (it then roughly equals to the inverse healing length of the BEC).
All other parameters are the same as in Fig.~\ref{fig:Low}.
Comparing the results of free fermion approximation in Figs.~\ref{fig:High}(a) and \ref{fig:High}(c) with those of dressed fermion in Figs.~\ref{fig:High}(b) and \ref{fig:High}(d), one sees that they are only slightly different.
In particular, MDFC of $\lambda = 0.5$ in Fig.~\ref{fig:High}(d) does not show any significant change as compared to (c), which is to be contrasted with the picture in Fig.~\ref{fig:Low}(d).

The reason for the difference in MDFC is that in Fig.~\ref{fig:Low}(d) the Fermi level is in the vicinity of the SWS, i.~e. in the vicinity of the crossing point of the fermion and boson dispersion. This is not the case for the data of Fig.~\ref{fig:High}(d), where the Fermi level lies far above the avoided crossing region. This crossover phenomenon is very interesting and we expect it to be observable experimentally.

\subsection{Vertex correction and the SWS}
\label{sec:vertex}

The calculations in the previous sections are performed with the lowest order self-energy diagram.
Its predictive power decreases significantly with growing interaction strength. In order to access the regime of intermediate to strong interactions one needs to use more efficient approximation schemes. In Ref.~\cite{Ji2014} it was shown that RPA cannot recover the SWS, hence an alternative one is necessary.
In this section we shall take into account the contribution from a two-phonon process, and compare the diagrammatic calculations with the numerically exact path integral QMC results \cite{Pollet2012}.
It allows for a calculation of the dynamic correlation function, from which the spectral function can be safely extracted.
The technical details of QMC simulations are described in Ref.~\cite{Ji2014}.
In order to improve the spectral resolution, in the present work we have increased the size of 1D system to 41 states.
Since QMC simulation cannot reach very low temperatures, in this section we shall fix the inverse temperature at $\beta = 10$ for both QMC and diagrammatic calculations.

The Feynman diagrams of two-phonon processes are depicted in Figs.~\ref{fig:Diagram}(d) and \ref{fig:Diagram}(e).
Panel (d) can be understood as a self-energy of the lowest order with a vertex correction to the fermion self-energy, while the process on panel (e) is known as the ring diagram.
The latter contributes to a screening effect on the inter-particle interactions, which is important for the high density fermionic systems such as the electron gases in metals.
In the present study, since we are mainly concerned with the cases of low fermion concentration, we shall neglect diagram (e) in the discussion below.
As for the diagram (d), the frequency summations can be performed in the same way as we have done previously.
After some algebra, the second order self-energy with the vertex corrections is obtained to be given by
\begin{widetext}
\begin{eqnarray}
\label{eq:vertex}
& \Sigma^{(2)} & ({\bf k}, i \varepsilon_n) = {1 \over V^2} \sum_{{\bf q}_1, {\bf q}_2} V_{{\bf q}_1}^2 V_{{\bf q}_2}^2
    \left\{ - {1 \over \beta} \sum_{i \omega_n}
    {\mathcal G}_0 ({\bf k} + {\bf q}_1, i \varepsilon_n + i \omega_n)
    {\mathcal D}_0 ({\bf q}_1, i \omega_n)
    \right. \nonumber \\
& & \,\,\,\,\,  \,\,\,\,\,\,  \,\,\,\,\,  \,\,\,\,\,  \,\,\,\,\,  \,\,
    \times \left. \left[ - {1 \over \beta} \sum_{i \nu_n}
    {\mathcal G}_0 ({\bf k} + {\bf q}_1 + {\bf q}_2, i \varepsilon_n + i \omega_n + i \nu_n)
    {\mathcal D}_0 ({\bf q}_2, i \nu_n)
    {\mathcal G}_0 ({\bf k} + {\bf q}_2, i \varepsilon_n + i \nu_n) \right] \right\}
    \nonumber \\
& = & {1 \over V^2} \sum_{{\bf q}_1, {\bf q}_2} V_{{\bf q}_1}^2 V_{{\bf q}_2}^2 \sum_{r,s = \pm 1}
    r s \left\{ \frac{1}{i \varepsilon_n + r \omega_{{\bf q}_1} + s \omega_{{\bf q}_2} - \xi_{{\bf k}+{\bf q}_1+{\bf q}_2}}
    \left[ \frac{N_B \left( r \omega_{{\bf q}_1} \right) N_B \left( s \omega_{{\bf q}_2} \right) }
    { \left( i \varepsilon_n - \xi_{{\bf k}+{\bf q}_1} + r \omega_{{\bf q}_1} \right)
    \left( i \varepsilon_n - \xi_{{\bf k}+{\bf q}_2} + s \omega_{{\bf q}_2} \right)}
    \right. \right . \nonumber \\
& & - \left. \frac{N_B \left( r \omega_{{\bf q}_1} \right) n_F \left( \xi_{{\bf k}+{\bf q}_1+{\bf q}_2} \right) }
    { \left( \xi_{{\bf k}+{\bf q}_2} - \xi_{{\bf k}+{\bf q}_1+{\bf q}_2} + r \omega_{{\bf q}_1} \right)
    \left( i \varepsilon_n - \xi_{{\bf k}+{\bf q}_1} + r \omega_{{\bf q}_1} \right)}
    + \frac{N_B \left( -s \omega_{{\bf q}_2} \right) n_F \left( \xi_{{\bf k}+{\bf q}_1+{\bf q}_2} \right) }
    { \left( \xi_{{\bf k}+{\bf q}_1} - \xi_{{\bf k}+{\bf q}_1+{\bf q}_2} + s \omega_{{\bf q}_2} \right)
    \left( i \varepsilon_n - \xi_{{\bf k}+{\bf q}_2} + s \omega_{{\bf q}_2} \right)} \right]
    \nonumber \\
& & + \frac{1}{i \varepsilon_n + \xi_{{\bf k}+{\bf q}_1+{\bf q}_2} - \xi_{{\bf k}+{\bf q}_1} - \xi_{{\bf k}+{\bf q}_2}}
    \left[ \frac{n_F \left( \xi_{{\bf k}+{\bf q}_1+{\bf q}_2} \right) n_F \left( \xi_{{\bf k}+{\bf q}_1} \right) }
    { \left( i \varepsilon_n - \xi_{{\bf k}+{\bf q}_1} + r \omega_{{\bf q}_1} \right)
    \left( \xi_{{\bf k}+{\bf q}_1} - \xi_{{\bf k}+{\bf q}_1+{\bf q}_2} + s \omega_{{\bf q}_2} \right) }
    \right. \nonumber \\
& & - \left. \frac{n_F \left( \xi_{{\bf k}+{\bf q}_1+{\bf q}_2} \right) n_F \left( - \xi_{{\bf k}+{\bf q}_2} \right) }
    { \left( i \varepsilon_n - \xi_{{\bf k}+{\bf q}_2} + s \omega_{{\bf q}_2} \right)
    \left( \xi_{{\bf k}+{\bf q}_2} - \xi_{{\bf k}+{\bf q}_1+{\bf q}_2} + r \omega_{{\bf q}_1} \right) }
    - \frac{n_F \left( \xi_{{\bf k}+{\bf q}_1} \right) n_F \left(- \xi_{{\bf k}+{\bf q}_2} \right) }
    { \left( i \varepsilon_n - \xi_{{\bf k}+{\bf q}_1} + r \omega_{{\bf q}_1} \right)
    \left( i \varepsilon_n - \xi_{{\bf k}+{\bf q}_2} + s \omega_{{\bf q}_2 } \right) } \right]
    \nonumber \\
& & + \left. \frac{1}{ \left( i \varepsilon_n - \xi_{{\bf k}+{\bf q}_1} + r \omega_{{\bf q}_1} \right)
    \left( i \varepsilon_n - \xi_{{\bf k}+{\bf q}_2} + s \omega_{{\bf q}_2 } \right) }
    \left[ \frac{ N_B \left( r \omega_{{\bf q}_1} \right) n_F \left( \xi_{{\bf k}+{\bf q}_2} \right) }
    {\xi_{{\bf k}+{\bf q}_2} - \xi_{{\bf k}+{\bf q}_1+{\bf q}_2}+ r \omega_{{\bf q}_1}}
    + \frac{ N_B \left( s \omega_{{\bf q}_2} \right) n_F \left( \xi_{{\bf k}+{\bf q}_1} \right) }
    {\xi_{{\bf k}+{\bf q}_1} - \xi_{{\bf k}+{\bf q}_1+{\bf q}_2}+ s \omega_{{\bf q}_2}}
    \right] \right\} \, .
\end{eqnarray}
\end{widetext}

Eq.~(\ref{eq:vertex}) can only be evaluated numerically.
Before we embark on that, we would like to point out some technical issues.
We perform the numerics in several steps.
We first compute the self-energy up to the two-phonon processes, $\Sigma = \Sigma^{(1)} + \Sigma^{(2)}$. Next we derive the spectral function Eq.~(\ref{eq:hf3}).
 Here one has to be cautious with the numerical instability related with $\mbox{Im} \Sigma^R$.
If $\mbox{Im} \Sigma^R$ becomes zero for some $\bf k$ and $\varepsilon$, then in Eq.~(\ref{eq:hf3}) it should be replaced in both numerators and denominators by $(\mbox{Im} \Sigma^R - 0^+)$. (See also the discussion in Sec.~\ref{sec:pseudo-gap}.)
In the last stage of numerics, the Bragg spectra are computed with the help of Eq.~(\ref{eq:hf2}).
In this procedure, the most time-consuming step is the multi-dimensional integration of $\Sigma^{(2)}$.
Moreover, in this step particular attention needs to be paid while performing the numerical analytic continuation
$i \varepsilon_n \to \varepsilon + i 0^+$ on $\Sigma^{(2)}$.
Here $0^+$ is roughly the numerical resolution of energy, and we fix it at 0.05 in our calculations (see Appendix~\ref{sec:app_recipe} for a detailed discussion, $0^+$ is measured in units of $\hbar^2/m_I \xi^2$).

\begin{figure}[htbp]
\includegraphics[width=0.45\textwidth]{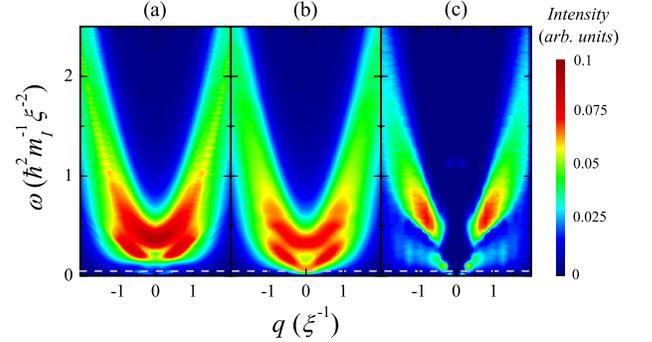}
\caption{
(Color online) Bragg spectrum of fermion impurities immersed in 1D BEC at Fermi momentum $k_F \simeq 0.1 \pi \xi^{-1}$, coupling constant $\lambda = 0.5$ and inverse temperature $\beta = 10$.
Calculation of panel (a) is based on the lowest order self-energy of fermion, using Eqs.~\eqref{eq:hf2}, \eqref{eq:hf3} and \eqref{eq:hf4}.
Panel (b) takes into account the vertex corrections from Eq.~\eqref{eq:vertex}. Panel (c) shows the results of a QMC simulation.
The horizontal gray dashed line denotes the Fermi energy.
\label{fig:B1D}}
\end{figure}

We concentrate here on Bragg spectrum as the most relevant quantity from the experimental point of view.
The related numerical results are presented in Fig.~\ref{fig:B1D}, where we compare the Bragg spectra of fermionic impurities of a 1D system computed with different methods. One can clearly see, that while the lowest order self-energy (panel (a)) is not able to resolve the gap between the upper and lower branches, taking into account the vertex corrections makes them clearly distinquishable, see panel (b) of Fig.~\ref{fig:B1D}. 
In panel (c), the QMC results are presented, where distinctive upper and lower branches are also observed, consistent with the diagrammatic calculations.
It is clearly seen in Fig.~\ref{fig:B1D} that taking the vertex corrections into account yields a better approximation of the QMC data.
We stress again, that this is in strong contrast to the RPA procedure, which fails to reproduce this feature.
In panel (c), one notices that the upper branches are separated near $k = 0$.
This is an artifact, probably suggesting that a system of 41 states in QMC is still not sufficient to fully eliminate the finite size effect.

\begin{figure}[htbp]
\includegraphics[width=0.4\textwidth]{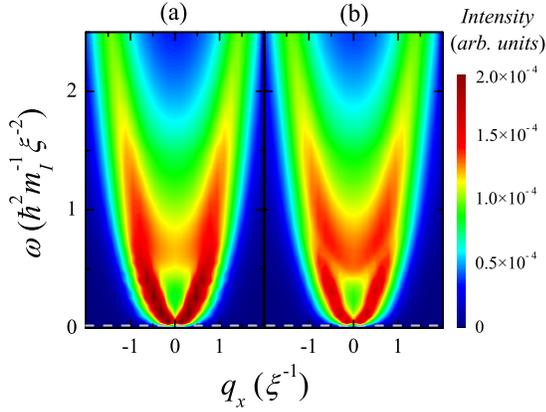}
\caption{(Color online) Bragg spectra of fermions in 2D BEC along the $q_x = q_y$ direction in momentum space for 
$\lambda = 1.2$
and inverse temperature $\beta = 10$.
The radius of the Fermi circle is $k_F \simeq 0.02 \pi \xi^{-1}$.
Calculations of panel (a) are based on the lowest order self-energy.
Panel (b) takes into account vertex corrections.
The horizontal gray dashed line denotes the Fermi energy.
}
\label{fig:B2D}
\end{figure}

It is instructive to test the robustness of our approximation scheme for systems of higher dimensions.
In 2D our diagramatic calculations indicate that the SWS exists at low fermion density or/and large coupling strength.
In Fig.~\ref{fig:B2D}, we plot the Bragg spectrum for a 2D model with $\lambda = 1.2$ and $\beta = 10$ on a simple quadratic lattice.
The Fermi surface is a circle with radius $k_F = 0.02 \pi$, which corresponds to low impurity density.
Here panel (a) depicts the Bragg spectra along the $q_x = q_y$ direction computed by the lowest-order Feynman diagram, and panel (b) displays the diagrammatic results including vertex correction along the same direction.
The effect of the vertex corrections is clearly distinguishable.
Panel (a) shows only very weak band mixing, while panel (b) exposes a more pronounced separation between the upper and lower branches.
We have also performed QMC simulations on 2D clusters comprising up to $9 \times 9$ states.
The results confirm the existence of a SWS in 2D.
(The system size for QMC in 2D is, however, not large enough to show continuous dispersion relations. Therefore we abstain from presenting the data.)
From the experimental point of view, 2D BECs are rather routinely realizable in ultracold atomic setups. Therefore, we expect that the SWS predicted above can be observed via measurement of Bragg spectra in ultracold mixtures. 

Given the pronounced weakening of the 2D SWS in comparison to that in 1D, we speculate that in 3D it might be very weak or even vanish altogether. An ultimate answer to that question would require a more detailed analysis and is an avenue for further research.

\section{Conclusions}
\label{sec:conclusion}

In this work we apply Feynman diagrammatics to study an ultracold fermion-boson gas mixture in the Bogoliubov regime. In such setups the energy dispersion of fermions is significantly modified by the fermion-boson interaction which leads to the spectral weight suppression (SWS) in the small momentum region. This phenomenon was previously found in QMC simulations. We have constructed a diagrammatic approach which consistently recovers the SWS and successfully applied it to compute three experimentally accessible key quantities: (i) single-particle spectral function; (ii) the Bragg spectrum; and (iii) the momentum dependent force autocorrelation function. We have shown that at low impurity densities one reliably observes the SWS feature in all three observables.
This is in contrast to the previously used RPA method, which is not able to recover the SWS. The results of our method are qualitatively confirmed by the QMC simulations and we expect that our predictions can be tested in the state-of-the-art experiments such as those presented in e.~g. Refs.~\cite{Schuster2012, Scelle2013} (fermionic $^{6}$Li impurities in a  $^{23}$Na condensate) very soon.

\begin{acknowledgments}
The authors thank Jonas Vlietinck, Wim Casteels, Sergei Klimin, Jacques Tempere, Jozef Devreese, Eddy Timmermans, and Markus Oberthaler for enlightening discussions. AK is supported by the Heisenberg Programme of the Deutsche Forschungsgemeinschaft (Germany) under Grant No. KO 2235/5-1.
\end{acknowledgments}

\appendix

\section{Drag force on fermionic impurities interacting with BEC}
\label{sec:app_force}

Here we derive the expression for the drag force acting on a fermionic impurity when it is immersed into a BEC.
We start from the original interacting boson-impurity Hamiltonian,
\begin{eqnarray}
\label{eq:H_BI}
H & = & H_B + H_{BB} + H_I + H_{IB} \, ,
                          \nonumber  \\                        \nonumber
H_B & = & \int d {\bf x} \Psi^*_B ({\bf x}) \left[ \epsilon_B ({\bf k}) -\mu_B \right] \Psi_B ({\bf x}) \, ,
    \\
H_{BB} & = & {1 \over 2} \int d {\bf x} d {\bf x'} \Psi^*_B ({\bf x}) \Psi^*_B ({\bf x'}) V_{BB} ({\bf x-x'})
    \nonumber \\                        \nonumber
& & \times \Psi_B ({\bf x'}) \Psi_B ({\bf x}) \, ,
    \\                        \nonumber
H_I & = & \int d {\bf x} \Psi^*_I ({\bf x}) \left[ \epsilon_I ({\bf p}) -\mu_I \right] \Psi_I ({\bf x}) \, ,
    \\                        \nonumber
H_{IB} & = & \int d {\bf x} d {\bf x'} \Psi^*_B ({\bf x}) \Psi_B ({\bf x}) V_{IB} ({\bf x-x'})
    \nonumber \\                        \nonumber
& & \times \Psi^*_I ({\bf x'}) \Psi_I ({\bf x'}) \, ,
\end{eqnarray}
where $\Psi_B ({\bf x})$ and $\Psi_I ({\bf x'})$ are the field operators of the boson and impurity, respectively, $\epsilon_B ({\bf k}) = \hbar^2 k^2 / (2 m_B)$ and $\epsilon_I ({\bf p}) = \hbar^2 p^2 / (2 m_I)$ are the dispersions of the boson and impurity, and $\mu_B$ ($\mu_I$) is the chemical potential of the boson (impurity).
The boson-boson contact potential has a form of $\delta$ function, $V_{BB} ({\bf x-x'}) = g_{BB} \delta ({\bf x-x'})$. The same is assumed for the impurity-boson interaction, $V_{IB} ({\bf x-x'}) = g_{IB} \delta ({\bf x-x'})$, with $g_{BB}$ and $g_{IB}$ being the scattering strengths.
The drag force acting on the impurity is given by \cite{Astrakharchik2004},
\begin{eqnarray}
\label{eq:app_f1}
{\bf F}_I & = & -  \int  d {\bf x} d {\bf x'} |\Psi_B ({\bf x})|^2 |\Psi_I ({\bf x'})|^2 \, \nabla_{\bf x'} V_{IB} (\bf{x-x'})
    \nonumber \\
& = & g_{IB} \int d {\bf x} |\Psi_B ({\bf x})|^2 \left[ \nabla_{\bf x'} |\Psi_I ({\bf x'})|^2 \right]_{\bf x'=x} .
\end{eqnarray}
To get Eq.~(\ref{eq:app_f1}), a partial integration has been applied ($\bf x'$ refers to the coordinate of the impurity).
Next we make a transition to the second quantization representation,
\begin{eqnarray}
\label{eq:app_f2}
\Psi_B ({\bf x}) = {1 \over \sqrt{V}} \sum_{\bf q} A_{\bf q} e^{i {\bf q} \cdot {\bf x}} \, ,
    \\
\label{eq:app_f3}
\Psi_I ({\bf x'}) = {1 \over \sqrt{V}} \sum_{\bf p} c_{\bf p} e^{i {\bf p} \cdot {\bf x'}} \, .
\end{eqnarray}
Substituting Eqs.~(\ref{eq:app_f2}) and (\ref{eq:app_f3}) into (\ref{eq:app_f1}), we find that the force operator becomes
\begin{eqnarray}
\label{eq:app_f4}
{\bf F}_I = - {i g_{IB} \over V} \sum_{\bf k,p,q} {\bf q} A^{\dag}_{\bf p-q} A_{\bf p} c^{\dag}_{\bf k} c_{\bf k-q} \, .
\end{eqnarray}
Under the Bogoliubov approximation, the boson number operator can be expressed as a small fluctuation superimposed on the condensate, i.~e. $A \rightarrow \sqrt{N_0} + a$ with $N_0 = n_0 V$.
Keeping the leading order nontrivial term, we rewrite the force operator as
\begin{eqnarray}
\label{eq:app_f5}                        \nonumber
{\bf F}_I = - i g_{IB} \sqrt{n_0 \over V} \sum_{\bf k,q} {\bf q} \left( a_{\bf q} + a^{\dag}_{\bf -q} \right) c^{\dag}_{\bf k} c_{\bf k-q} \, .
\end{eqnarray}
In order to make the notation consistent with that of the Fr\"{o}hlich model for BEC polaron, a Bogoliubov transformation needs to be applied \cite{Pitaevskii2003},
\begin{eqnarray}
\label{eq:app_f6}                        \nonumber
\left\{
\begin{array}{l}
a_{\bf q} = u_{\bf q} b_{\bf q} - v_{\bf q} b^{\dag}_{\bf -q}
    \\
a^{\dag}_{\bf q} = u_{\bf q} b^{\dag}_{\bf q} - v_{\bf q} b_{\bf -q}
\end{array}
\right.
\end{eqnarray}
with the coefficients satisfying
\begin{eqnarray}                        \nonumber
\label{eq:app_7}
u_{\bf q} - v_{\bf q} = {V_{\bf q} \over g_{IB} \sqrt{n_0}} \, .
\end{eqnarray}
Then we finally obtain the force operator as
\begin{eqnarray}
\label{eq:app_f5}                        \nonumber
{\bf F}_I = - {i \over \sqrt{V}} \sum_{\bf k,q} {\bf q} V_{\bf q} \left( b_{\bf q} + b^{\dag}_{\bf -q} \right) c^{\dag}_{\bf k} c_{\bf k-q} \, .
\end{eqnarray}

\section{RPA calculation of the Bragg spectrum of fermion-boson mixture}
\label{sec:app_rpa}

\begin{figure}[htbp]
\includegraphics[width=0.45\textwidth]{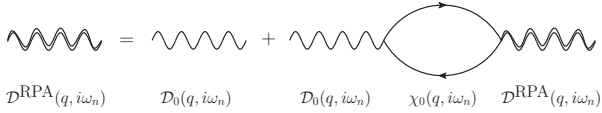}
\caption{Phonon Dyson equation in random phase approximation.
The straight line stands for a naked fermion, and the single (double) wavy line for a naked (dressed) phonon.}
\label{fig:App_RPA}
\end{figure}

The details of RPA calculation have been elaborated in an earlier paper \cite{Ji2014}.
Here we only outline the main procedure together with the representative result for a 1D system.
Instead of working on the Bragg spectrum $\mathcal{R} (q, \omega)$ directly, we start with calculating the optical conductivity,
\begin{eqnarray}
\label{eq:absorb}
\mbox{Re} [\sigma (q, \omega)] = -{1 \over \omega} \mbox{Im} \, \Pi^R (q, \omega) \, , 
\end{eqnarray}
because these two quantities are closely related to each other \cite{Casteels2011a},
\begin{eqnarray}
\label{eq:fund_relation}
\mathcal{R} (q, \omega) = {q^2 \over \pi \omega} \mbox{Re} [\sigma (q, \omega)] \, .
\end{eqnarray}
$\Pi^R (q, \omega)$ in Eq.~(\ref{eq:absorb}) is the the current-current correlation function.
In the Matsubara representation, we have
\begin{eqnarray}
\label{eq:cur_cur}
\Pi (q, i \omega_n) = -{1 \over V} \int_0^{\beta} d \tau e^{i \omega_n \tau}
\langle T_\tau j^{\dag} (q, \tau) j (q,0) \rangle ,
\end{eqnarray}
where
\begin{eqnarray}
\label{eq:current}
& & j (q) = {1 \over m_I} \sum_k \left(k + {q \over 2} \right) a^\dag_{k+q} a_k \, ,
\end{eqnarray}
is the current operator for a fermion of momentum $q$.
After applying two partial integrations to the right hand side of Eq.~(\ref{eq:cur_cur}), we obtain
\begin{widetext}
\begin{eqnarray}
\label{eq:cur_cur3}
& & \Pi (q, i \omega_n) = \frac{q^2}{(i \omega_n)^2 m^3_I V} \sum_k \left(3k^2 + {q^2 \over 4} \right) \langle a_k^{\dag} a_k \rangle
- \frac{1}{(i \omega_n)^2 m^2_I V} \sum_{q'} q' (q + q') V_{q'}^* \langle B_{q'} \rho^{\dag} (q') \rangle
- \frac{1}{(i \omega_n)^2 m^2_I V} \int_0^{\beta} d\tau e^{i \omega_n \tau}
\nonumber \\
& & \times \left[ \frac{q^2}{m^2_I} \sum_{kk'} \left( k + {q \over 2} \right)^2 \left( k' + {q \over 2} \right)^2
\langle T_\tau a_k^{\dag} (\tau) a_{k+q} (\tau) a_{k'+q}^{\dag} a_{k'} \rangle
- \frac{q}{m_I} \sum_{kq'} V_{q'} q' \left( k + {q \over 2} \right)^2
\langle T_\tau B_{q'}^{\dag} a_k^{\dag} (\tau) a_{k+q} (\tau) \rho (q+q') \rangle
\right.
\nonumber \\
& & \left. - \frac{q}{m_I} \sum_{kq'} V_{q'}^* q' \left( k + {q \over 2} \right)^2
\langle T_\tau B_{q'} \rho^{\dag} (q+q', \tau) a_{k+q}^{\dag} a_k \rangle
+ \sum_{q'q''} V_{q'}^* V_{q''} q' q''
\langle T_\tau B_{q'} (\tau) B_{q''}^{\dag} \rho^{\dag} (q+q', \tau) \rho (q+q'') \rangle \ \right] \, , 
\end{eqnarray}
\end{widetext}
As noted in the early studies on the optical conductivity \cite{Mahan2000}, the advantage of the partial integration scheme is that it can identify the electron-phonon coupling effect, e.~g. the formation of the Fr\"{o}hlich polaron, in its leading order terms.
On the contrary, in a direct perturbative expansion of the current autocorrelation function such effects are usually hidden in the higher order terms.
Following the conventional treatment, let us keep the leading order contributions in Eq.~(\ref{eq:cur_cur3}), i.~e. the first and the last terms in the square brackets.
After performing an analytical continuation of these two terms, i.~e. imposing $i \omega_n \rightarrow \omega + i 0^+$, and recalling the relation in Eq.~(\ref{eq:fund_relation}), we get an expression for the Bragg spectrum,
\begin{eqnarray}
\label{eq:Bragg_RPA} 
\mathcal{R} (q,\omega) = & & {m_I \over 2 \pi |q|} n_F (\epsilon_p)|_{p={m_I \omega \over q} - {q \over 2}} \left[ 1 - n_F (\epsilon_p)|_{p={m_I \omega \over q} + {q \over 2}} \right]
\nonumber \\
& & \times \left( 1 - e^{-2 \beta m_I \omega} \right)
    + \frac{q^2}{\pi m_I \omega^4 V^2} \sum_{s=\pm} \sum_{p q'} s |V_{q'}|^2
\nonumber \\
& & \times {q'^2 \over |q+q'|} n_F (\epsilon_p) [ n_B (s \epsilon_{p+sq+sq'} - s \epsilon_p)
\nonumber \\
& & -n_B (s \epsilon_{p+sq+sq'} - s \epsilon_p - \omega) ]
\nonumber \\
& & \times \mbox{Im} D^R (q', s \epsilon_{p+sq+sq'} - s \epsilon_p - \omega) \, . 
\end{eqnarray}
We then proceed with the calculation under RPA.
The essential idea is to interpret the phonon as a quasiparticle modulated by the fermion-boson interaction and still treating the fermion as being free.
The phonon is hence dressed by the virtual excitations and acquires a self-energy, as schematically shown in Fig.~\ref{fig:App_RPA}.
The phonon Green's function is determined by a Dyson equation,
\begin{eqnarray}                        \nonumber
\mathcal{D}^{\mathrm{RPA}} (q, i \omega_n) = \frac{\mathcal{D}_0 (q, i \omega_n)}{1 - V^2_q \mathcal{D}_0 (q, i \omega_n) \chi_0 (q, i \omega_n)} \, ,
\end{eqnarray}
where $\mathcal{D}_0 (q, i \omega_n)$ and $\chi_0 (q, i \omega_n)$ are the free phonon Green's function and free fermion density-density correlation function, respectively.

\begin{figure}[htbp]
\includegraphics[width=0.45\textwidth]{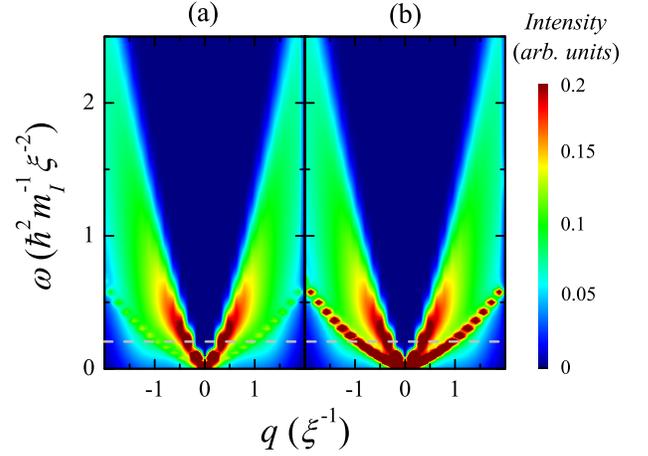}
\caption{(Color online)
Bragg spectra calculated with RPA for different coupling strengths: (a) $\lambda = 0.36$, and (b) $\lambda = 0.72$.
The calculations are performed for 1D systems at $\beta = 10$ with their Fermi momenta at $k_F = 0.2 \pi \xi^{-1}$.
The horizontal gray dashed line denotes the Fermi energy.
}
\label{fig:App_Bragg}
\end{figure}

The numerical results on Bragg spectra are shown in Fig.~\ref{fig:App_Bragg} for two different coupling strengths.
It is evident that some peaks are enhanced for increasing $\lambda$.
They come from the second term of Eq.~(\ref{eq:Bragg_RPA}), reflecting the `optical' excitations of Bogoliubov phonon modes.
One can see that the resultant spectrum is consistent with the dispersion relation of the Bogoliubov mode schematically depicted in the inset of Fig.~\ref{fig:A}(a).
In addition, there is a broad continuum in the background seemingly independent of $\lambda$.
This plateau comes from the excitations of fermions across the Fermi level, i.~e. the excitonic excitations, described by the first term of Eq.~(\ref{eq:Bragg_RPA}).
Despite these characteristic components, ones immediately notices that a main difference of Fig.~\ref{fig:App_Bragg} from Fig.~\ref{fig:B1D} is the absence of the SWS.
There is no relative shift of the spectral components even at large $\lambda$, which is a fundamental drawback of the RPA approach.

\section{Numerical analytic continuation of vortex correction}
\label{sec:app_recipe}

In the quantum many-body theory, the extraction of a dynamical quantity of real frequency (or real time) from its Matsubara counterpart is known as analytic continuation (AC).
By its original definition, AC can be established by a substitution $i \varepsilon_n \rightarrow \varepsilon + i 0^+$ if the functional form of the physical quantity is already known, where $0^+$ means an infinitesimal positive number which is necessary to enforce the causality.
Without knowing the analytical expression, AC of the physical quantity still can be performed by the numerical methods but it becomes notoriously difficult because of its ill-posed nature.
The results are sensitive to various numerical errors which can come from, for example, a cutoff in the summation over Matsubara frequencies, or the numerical noise inherent to QMC simulations.
In our calculation, thanks to the analytical expressions for the self-energy obtained in Eqs.~(\ref{eq:hf4}) and (\ref{eq:vertex}), we are able to accomplish the AC with a good accuracy through the simple conversion $i \varepsilon_n \rightarrow \varepsilon + i 0^+$. 

In the operation of AC, Eq.~(\ref{eq:hf4}) is rather easy to handle, while vertex correction Eq.~(\ref{eq:vertex}) still requires special attention and techniques to avoid unphysical artifacts.
To see the origin of the difficulty, one can decompose the products in Eq.~(\ref{eq:vertex}) and rewrite it as a summation: 
\begin{eqnarray}
\label{eq:ac1}
\Sigma^{(2)} (i \varepsilon_n) = \sum_j {C_j \over i \varepsilon_n + E_j} \, ,
\end{eqnarray}
where $C_j$ and $E_j$ represent some constants.
Since Eq.~(\ref{eq:ac1}) has the same form as Eq.~(\ref{eq:hf6}), the imaginary part of the retarded self-energy can be separated as in Eq.~(\ref{eq:hf7}), which gives
\begin{eqnarray}
\label{eq:ac2}
\mbox{Im} \Sigma^{(2) R} (\varepsilon) = - \pi \sum_j C_j \delta (\varepsilon + E_j) \, ,
\end{eqnarray}
with $\delta (\varepsilon)$ the Dirac delta function.
However, it should be noted that there is essential difference between Eqs.~(\ref{eq:hf6}) and (\ref{eq:ac1}) --  all $C_j$ in $\Sigma^{(1) R}$ are non-negative, while $\Sigma^{(2) R}$ may have some negative $C_j$.
A direct consequence of negative $C_j$ is a possible violation of the positivity of spectral function $\mathcal{A} (\varepsilon)$, as can be seen from its relation with $\mbox{Im} \Sigma^R (\varepsilon)$ in Eq.~(\ref{eq:hf3}). This is an artefact of the approximation. In the full perturbation series a mutual term cancellation must take place, so that
the overall $\mathcal{A} (\varepsilon)$ would remain positive.
While under certain truncation, negative unphysical $\mathcal{A} (\varepsilon)$ can appear in some cases.

However, it turns out that the spectral weight negativity has almost no influence on our results. It is less pronounced the larger values we choose for the infinitesimal $0^+$. While a value around $\simeq0.01$ generates noticeable artefacts, $0^+$ of the order $0.05$ makes them virtually invisible. On the other hand the latter value is of the order of the numerical energy resolution of the Monte Carlo data, rendering values of $0^+$ less then $0.05$ meaningless. That is why we have used this value for $0^+$ throughout.

\bibliography{BECP_Pseudogap_v1}

\end{document}